\documentclass{emulateapj}

\begin{document}

\title{Spectral State Transitions of the Ultraluminous X-ray Sources X-1 and X-2 in NGC 1313}

\author{Hua Feng and Philip Kaaret}

\affil{Department of Physics and Astronomy, The University of Iowa, Van
Allen Hall, Iowa City, IA 52242; hua-feng@uiowa.edu}

\shortauthors{Feng and Kaaret}
\shorttitle{ULXs in NGC 1313}
\begin{abstract}
X-ray spectral state transitions are a key signature of black hole X-ray binaries and follow a well-defined pattern.  We examined 12 {\it XMM-Newton} observations of the nearby spiral galaxy NGC 1313, which harbors two compact ultraluminous X-ray sources (ULXs), X-1 and X-2, in order to determine if the state transitions in ULXs follow the same pattern.  For both sources, the spectra were adequately fitted by an absorbed power-law with the addition of a low temperature ($kT = 0.1 \sim 0.3$ keV) disk blackbody component required in 6 of the 12 observations.  As the X-ray luminosity of X-1 increases to a maximum at $3\times10^{40}$ ergs s$^{-1}$, the power-law photon index softens to 2.5--3.0. This behavior is similar to the canonical spectral state transitions in Galactic black hole binaries, but the source never enters the {\it high/soft} or {\it thermal dominant} state and instead enters the {\it steep power-law} state at high luminosities.  X-2 has the opposite behavior and appears to be in the {\it hard} state, with a photon index of $\Gamma = 1.7 - 2.0$ at high luminosity, but can soften to $\Gamma = 2.5$ at the lower luminosities.
\end{abstract}


\section{Introduction}

Ultraluminous X-ray sources (ULXs) are point-like, non-nuclear X-ray sources with luminosities above the Eddington limit of a 10 $M_\sun$ black hole \citep[$2\times10^{39}$ ergs s$^{-1}$, for review see][]{fab06}.  Many ULXs show strong variability and are believed to be black hole binaries.  Their X-ray luminosities \citep{col99,kaa01}, thermal disk emissions \citep{kaa03}, variation time scales \citep{str03,dew06} and surrounding emission line nebulae \citep{kaa04b} suggest that they might have masses of 20--10$^3$ $M_\sun$, falling into the class of intermediate mass black holes (IMBHs).  However, if the emission is beamed \citep{kin01,kor02} or exceeds the Eddington limit \citep{beg02,ebi03}, some or all ULXs may be stellar-mass black holes. 

A comparison of the properties of ULXs with the well-known phenomenology of stellar-mass black hole X-ray binaries should help reveal the nature of ULXs.  Spectral state transitions are essential characteristics of black hole binaries \citep[for a review see][]{mcc06}.  Most X-ray binaries have been observed in both the {\it hard} and {\it high/soft} states.  Typically, the intrinsic energy spectrum in the {\it hard} state is described as a power-law with a photon index $\Gamma = 1.5 \sim 2.1$.  The energy spectrum in the {\it high/soft} consists of a thermal disk component and a power-law with a photon index $\Gamma = 2.1 \sim 4.8$.  Some sources also show the {\it quiescent} state with very low luminosity and the {\it steep power-law} state with $\Gamma>2.4$.  The {\it steep power-law} state differs from the {\it high/soft} state in that the disk is the dominant spectral component in the {\it high/soft} state while the power-law component is dominant in the {\it steep power-law} state.  The luminosity and spectral shape are correlated with higher luminosities corresponding to softer photon indexes, e.g., in Cyg X-1 \citep{wil06} and XTE J1550$-$564 \citep{kub04}, and to higher disk temperatures, e.g., in XTE J1550$-$564 \citep{kub04} and 4U 1630$-$47 \citep{tom05}.  However, in the {\it steep power-law} state, the disk luminosity remains constant or decreases while the disk temperature increases \citep{tom05,mcc06}.

If ULXs are binary systems similar to Galactic black hole binaries, then they should exhibit similar spectral state transitions, unless they are super-Eddington sources in which case the spectral state(s) may differ from the standard ones.  State transitions following the canonical {\it high/soft} versus {\it hard} pattern have been reported for two ULXs in IC 342 \citep{kub01} and Holmberg IX X-1 \citep{lap01}.  However, the temporal coverage for IC 342 was poor, only two observations, while the coverage for Holmberg IX X-1 was poor and had to be stitched together from several different observatories.  Multiple high-quality {\it XMM-Newton} spectra have been obtained only for a few sources: two very similar spectra for Holmberg IX X-1 \citep{wan04} and three observations of Holmberg II X-1 with one showing an unusual low/soft state \citep{dew04}.  Repeated {\it Chandra} observations of the Antennae revealed state transitions of several ULXs \citep{fab03}, but the photon counts were insufficient to permit spectral modeling and the source spectra were studied using hardness ratios -- which are insufficient to disentangle the disk versus power-law components of the spectra. 

The availability of 14 archival {\it XMM-Newton} observations of ULXs NGC 1313 X-1 and X-2 offers an opportunity to investigate state transitions of ULXs.  NGC 1313 is a spiral galaxy at a distance of 4.13$\pm$0.11 Mpc \citep{men02}.  Besides the supernova remnant SN 1978K, two ULXs, X-1 and X-2, are found to be associated with NGC 1313.  Both sources exhibit disk emission with temperatures lower than those found in stellar-mass black holes \citep{mil03}, suggesting they could be IMBHs.  In the luminosity versus temperature ($L$--$kT$) diagram of bright ULXs in nearby galaxies \citep{fen05}, they appear in the low temperature, high luminosity class, which contains sources most likely to be IMBH candidates.  The maximum luminosities of X-1 and X-2 exceed the Eddington limit for a 20 $M_\sun$ black hole by a factor of 10, which is difficult to produce in super-Eddington models. \citet{ram06} find a unique optical counterpart to NGC 1313 X-2 in {\it HST} images and suggest that the companion star is a B1-B2 giant.  NGC 1313 X-2 resides in an optical nebular supershell \citep{pak03}, whose kinetic energy is much larger than that could be produced by a typical supernova explosion.  The shell is either a hypernova remnant or a continually powered nebula - possibly by  an outflow from the ULX.  X-ray luminosities and spectral features, as well as optical counterparts, suggest these two ULXs are interesting IMBH candidates.  In the following, we present the details of the observations, our analysis, and our results on the spectral variation in \S\ref{sec:data} and discuss the nature of state transitions of these two ULXs in \S\ref{sec:diss}.

\section{Data Reduction}
\label{sec:data}

We analyzed 12 {\it XMM-Newton} observations of NGC 1313 from 2000 Oct to 2005 Feb (Table \ref{tab:spec}).  There are another two observations made on 2003-Sep-09 and 2003-Dec-27, which are not included because of high background contamination.  We reduced the ODF data to event files with SAS 6.5.0 and calibration files current as of May 2006. Data were selected from good time intervals, where there were no background flares, with the best quality data (FLAG=0) and screened with PATTERN$\le$12 for MOS and PATTERN$\le$4 for PN.  Source spectra were extracted in 32\arcsec\ radius circular regions.  Background regions were selected at the same CCD chip as the source and at a similar distance from the readout node.  We combined all available PN and MOS data from each individual observation for the spectral analysis, unless the source was located on or near a CCD gap in one or more of the instruments.  

Spectra were fitted with XSPEC 11.3.2.  First, we tried to fit data with an absorbed power-law model (wabs*powerlaw in XSPEC), and then added a multicolor disk blackbody model (wabs(diskbb$+$powerlaw) in XSPEC) to see if that improved the fitting.  If the disk blackbody component had a significance level above 99\% as evaluated using an F-Test, we accepted the second model as the best-fit model, otherwise, we adopted the single power-law model.  All the best-fit parameters and their errors at the 90\% confidence level are shown in Table~\ref{tab:spec} with the luminosity in the 0.3--10 keV band after correction for the absorption along the line of sight.  For observations best-fitted with a single power-law model, we also tried to add the disk blackbody component but that did not change the power-law index or luminosity significantly.  We attempted to use a single absorption column density to fit all the observations of each source, but found that adequate fits could not be obtained in several cases.  Thus, the absorption was allowed to vary.  We found that the normalizations for all three detectors were consistent within errors.

Other models such as a hot disk blackbody plus a soft power-law, a soft blackbody plus a hot disk blackbody, or a soft disk blackbody plus a hard Compton corona have been used to fit ULX spectra \citep{rob05,sto06}.  A prime motivation for such models has been spectral curvature at high energies.  We find no evidence for high energy spectral curvature for NGC 1313 X-1 or X-2.  We note that \citet{sto06} report a break at around 5 keV in X-1 in an analysis of only the high energy portion of the spectrum, however, this break is not reflected in their modeling of the broad band spectrum.  Furthermore, the hot disk blackbody plus a soft power-law model is unphysical \citep{sto06}, while the soft blackbody plus hot disk blackbody would require a very strongly super-Eddington source for X-1 or X-2.  We note that in our model of choice, the power-law component is a reasonable approximation to a more physical model such as a Compton corona if the optical depth is low.  Indeed, in applying a Comptonization model to NGC 1313 X-1 and X-2, \citet{sto06} find a low optical depth for X-1 and that the power-law model gives as good as fit as the Comptonization model for X-2 indicating that the optical is consistent with being low.  Since we find no strong evidence for spectral curvature at high energies, use of a Comptonization model is not justified.  Use of a soft disk blackbody plus power-law model provides the closest analogy to the spectral model commonly used for the classification of spectral states in Galactic black holes \citep{mcc06}.

The disk component was only detected in observations (1), (6), (8), (10)--(12) for X-1, with a fractional contribution of 0.27, 0.20, 0.32, 0.41, 0.20, and 0.27, respectively, to the total intrinsic luminosity in the 0.3--10 keV band.  Disk emission from X-2 was detected in observations (1), (3), (9)--(12) with a fractional flux of 0.21, 0.23, 0.40, 0.31, 0.24, and 0.63, respectively.  The disk fractional flux was calculated by setting the column density and power-law normalization to zero and finding the ratio of the resulting flux in the 0.3--10 keV band to the total unabsorbed flux in the same band.  We also calculated the fractional fluxes in the 0.1--10, 0.2--10, and 0.4--4 keV bands and got similar values.  We note that the disk luminosity obtained from the fitting is correlated with the absorption column depth.  In particularly, high disk luminosities tend to occur for high $n_{\rm H}$ values.  We consider the disk luminosity values to be good upper bounds, but some caution is required in their interpretation.

\begin{figure}[t]
\centering
\plotone{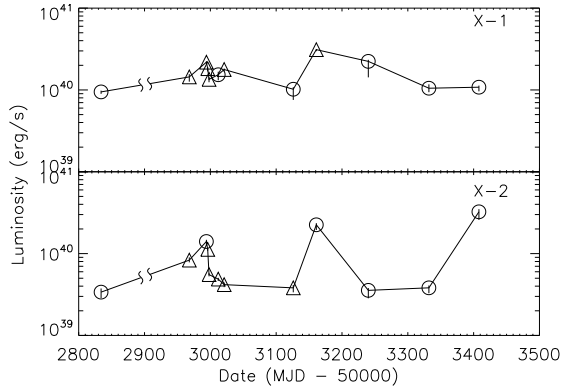}
\caption{Lightcurves of NGC 1313 X-1 (top) and X-2 (bottom). The first point is shifted 1000 days later to fit on the plot. Luminosities are taken from best-fit models: $\bigcirc$ -- disk blackbody plus power-law, $\bigtriangleup$ -- single power-law. 
\label{fig:lc}}
\end{figure}

X-ray lightcurves for NGC 1313 X-1 and X-2 are presented in Fig.~\ref{fig:lc}.  X-1 is variable by a factor of about 3, while X-2 is variable by a factor of about 10.

\begin{figure}[t]
\centering
\plotone{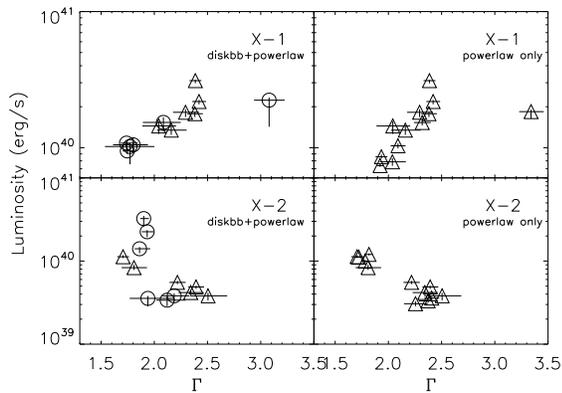}
\caption{Total X-ray luminosity in the 0.3--10 keV band versus power-law photon index for NGC 1313 X-1 (top) and X-2 (bottom).  For the panels on the left, the photon-index and luminosity are calculated from the best fitting model, either $\bigcirc$ -- disk blackbody plus power-law or $\bigtriangleup$ -- single power-law. For the panels on the right, the photon-index and luminosity are calculated from an absorbed power-law model for all observations.
\label{fig:lg}}
\end{figure}

A strong correlation was found between $\log(L_{\rm X})$ and photon index ($\Gamma$) for X-1, see the top-left panel of Fig.~\ref{fig:lg}, with higher luminosities found for softer spectra.  Including all 12 observations, the correlation coefficient is 0.82 with a chance probability of occurrence of $1.1\times10^{-3}$.  Excluding the rightmost point (observation 10) where the luminosity appears to saturate, the correlation coefficient is 0.91 and the chance probability is $8.8\times10^{-5}$.  We note that if we use a power-law only model to fit all the observations, then this correlation is essentially unchanged, see the top-right panel of Fig.~\ref{fig:lg}.  For X-2, the correlation follows the opposite trend with lower luminosities found for softer spectra, see the bottom-left panel of Fig.~\ref{fig:lg}. With the power-law only model, the points move into two clusters, one at high luminosity and hard photon index and the other at low luminosity and soft photon index, see the bottom-right panel of Fig.~\ref{fig:lg}.

We found some evidence for a correlation between the total ($L_{\rm total}$) and disk ($L_{\rm disk}$) luminosity for X-2.  However, the uncertainty in the disk luminosity warrants caution in the interpretation of this putative correlation.  We also examined correlations between $L_{\rm X}$, $L_{\rm disk}$, $L_{\rm powerlaw}$, $kT$, and $\Gamma$ for both sources, but no other significant correlations were found.

We searched for time variability for both sources in every observation. A careful screening procedure to avoid data gaps and background flares was applied.  Power spectra were calculated from continuous good time intervals and no significant variability was found. However, non-constant lightcurves with a significance level above 3$\sigma$ were found with a KS test in observations (3) and (9) for X-1 and observation (3) for X-2, respectively in a good time interval of 4.45, 7.23 and 4.45 ksec.  We note that the {\it XMM} EPIC CCD timing gaps have timescales from several to hundreds seconds, which can result in strong instrumental signals in the frequency range from mHz to Hz.  These instrumental effects can appear as artificial quasi-periodic oscillations or broken power-law features if adequate care is not taken in dealing with data gaps and background flares.  High quality temporal diagnostics, which could positively confirm 
the state identification, is, however, absent with these XMM data.

\section{Discussion}
\label{sec:diss}

The strong correlation between X-ray luminosity ($\log(L_{\rm X})$) and power-law photon index ($\Gamma$) for NGC 1313 X-1 is similar to the behavior found in Galactic black hole binaries, e.g., Cyg X-1 \citep{wil06}, and suggests that ULX NGC 1313 X-1 is a black hole X-ray binary.  The disk temperature in X-1 is constant within the uncertainties.  However, this is consistent with the behavior seen from stellar-mass black hole binaries because the range of the luminosity variation in X-1 is small.

In Galactic black hole X-ray binaries, over 75\% of the total flux in 2-20 keV band in the {\it high/soft} state arises from the disk emission component -- the {\it high/soft} state is a {\it thermal dominant} state \citep{mcc06}.  To compare the disk versus power-law luminosity ratios for ULXs, we choose to use the 0.3--10~keV band rather than the 2--20 keV band.  This is motivated because the the disk temperature in ULXs is a factor of 5--10 smaller than that in stellar-mass black holes.  We also compared the luminosity ratios in the 0.1--10 keV and in the 0.4--4 keV band, as suggested by the referee, and found they were very similar to those in the 0.3--10 keV band.

The fraction of total flux arising from the disk for X-1 is always below 50\%, even allowing for the uncertainties in the measurement of the disk luminosity.  Therefore, X-1 appears to never enter the {\it high/soft} state and is in either in the {\it hard} state or the {\it steep power-law} state. \citet{mcc06} define the {\it steep power-law} state by the presence of a power-law component with a photon index larger than 2.4.  In Fig.~\ref{fig:lg}, observation (10) with $\Gamma=3.08$ and a disk component of 41\% fractional flux, showing up as the rightmost point in Fig.~\ref{fig:lg}, is well consistent with the {\it steep power-law} state and there are observations ((3), (4), (7), and (9)) with a single power-law spectrum and $\Gamma = 2.3 \sim 2.4$ at the hard edge of the {\it steep power-law} state.  The non-detection of timing noise, except a little non-constant emission, and the absence of quasiperiodic oscillations are consistent with the {\it steep power-law} state when the power-law flux is greater than 50\% of the total \citep{mcc06}.  The luminosity of X-1 appears to saturates when the source enters the {\it steep power-law} state even though the photon index can soften significantly.  The {\it steep power-law} state is often associated with the highest luminosities seen from Galactic black hole X-ray binaries.  The five observations that we classify as the {\it steep power-law state} ((3), (4), (7), (9), (10)) all have luminosities near the maximum observed from X-1.  

The other observations of X-1 show photon index $\Gamma < 2.2$, reduced luminosity, and sometimes a disk component with a fractional flux of $\lesssim$30\%.  These properties are all consistent with classification as the {\it hard} state.  We note that relative small gap in luminosity between the {\it steep power-law} and the {\it hard} state is consistent with the behavior seen from some Galactic black hole binaries \citep{fen04,hom04,mcc06}.

The detection of state transitions from X-1 may help determine whether or not the emission is beamed.  Two models with anisotropic emissions, the mechanical beaming \citep{kin01} and relativistic beaming \citep{kor02}, have been raised to interpret the high luminosity of ULXs as radiations from stellar-mass black holes.  The same behavior of state transitions in X-1 as in stellar-mass black holes indicates the emission is from the accretion flow and not a relativistic jet.  The mechanical beaming model needs a near- or super-critical accretion rate in the binary system.  Whether or not such system can produce spectral state transition as observed in normal black hole binaries is unclear.  Other models involving super-Eddington accretion rates, like the advection-dominated optically thick disk \citep{ebi03} or the radiation-dominated accretion disk \citep{beg02}, must also address the spectral state transition behavior.  The transition to the {\it hard} state at luminosities only a factor of 3 below the maximum observed luminosity, as well as a maximum luminosity which exceeds the Eddington limit of a 20 $M_\sun$ black hole by a factor of 10, may be difficult to explain in such scenarios.  

The main physical distinction between the {\it steep power-law} state and the {\it high/soft} state (in most models) is the presence of a highly energized corona in the {\it steep power-law} state.  The absence of the {\it high/soft} state, or so called {\it thermal dominant} state, may suggest that NGC 1313 X-1 always has an energetically important corona.  \citet{goa06} and \citet{sto06} have suggested that ULXs have optical thick coronae, but the data for NGC 1313 X-1 suggest an optically thin corona.  The nature of the corona in X-ray binaries is still a mystery.  Some models suggest that magnetic fields play an important role in connecting the accretion disk and the corona through magnetorotational instability \citep[cf.][]{mil00} and the corona acts like a magnetic reservoir \citep{mer01}.  \citet{mer03} proposed a coupled magnetic disk-corona solution, in which the disk is stable when it is corona-dominated at high accretion rates. Most bright ULXs have spectra dominated by a soft power-law component \citep{fen05}, and are thought to be accreting from high-mass companion stars via Roche-lobe overflow \citep{liu04,kaa04a,ram06,kaa06a,kaa06b}.  Detailed modeling of accretion flows which produce powerful coronae at high accretion rates should be important in understanding ULXs and also in determining whether super-Eddington accretion occurs \citep{mer03}.  We note that, based on the spectral properties alone, we identify the state of Holmberg II X-1 reported by \citet{goa06} as the {\it steep power-law} state since the photon index is 2.6 and less than 15\% of the flux arises from the disk.  The low level of timing noise is consistent with the state definitions of \citet{mcc06}, but should be considered in more detail in relation to the timing properties of Galactic black hole binaries in the {\it steep power-law} state.

At luminosities above $6 \times 10^{39} \rm \, erg \, s^{-1}$, NGC 1313 X-2 appears to be in the {\it hard} state, with a photon index of $\Gamma = 1.7 - 2.0$, see the bottom panel of Fig.~\ref{fig:lg}.  At lower luminosities, the photon index can reach much softer values.  When fitting with the simple power law model, the points from two clusters with high luminosity associated with hard photon index and low luminosity with soft photon index. The high/hard vs low/soft behavior of X-2 is unlike that of typical Galactic black hole binaries, but a similar, unusual, low/soft state has been observed from GRS 1758-258 \citep{smi01}.  Galactic black holes in the {\it hard} state typically produce compact radio jets.  Scaling using the radio to X-ray flux ratio of GX 339-4 \citep{cor03}, the predicted radio flux density would be $\sim 5 \, \mu$Jy at 8.6~GHz and would be difficult to detect.  However, such a jet could power the optical nebula surrounding X-2. 

\acknowledgments 

We thank the anonymous referee for comments which improved our paper.  PK acknowledges partial support from a University of Iowa Faculty Scholar Award.


\begin{deluxetable}{llllp{0.1cm}lllp{0.1cm}p{0cm}l}
\tablecaption{Best-fit spectral parameters of NGC 1313 X-1 and X-2 in different observations\label{tab:spec}}
\tablehead{
\colhead{No.} & \colhead{obs. date} & 
\colhead{Instruments} & {Exposure} &
\colhead{$n_{\rm H}$} & \colhead{$\Gamma$} & \colhead{$kT$} &
\colhead{0.3--10 keV $L_{\rm X}$} & \colhead{0.3--10 keV $L_{\rm disk}$} & \colhead{$\chi^2/$dof}\\
\colhead{} & \colhead{} & 
\colhead{} & {(ksec)} &
\colhead{($10^{22}$ cm$^{-2}$)} & \colhead{} & \colhead{(keV)} &
\colhead{($10^{40}$ ergs s$^{-1}$)} & \colhead{($10^{40}$ ergs s$^{-1}$)} & \colhead{}
}
\startdata
&&&&&NGC 1313 X-1&&&&&\\
\noalign{\smallskip} \hline \noalign{\smallskip}
1 & 2000 Oct 17 & PN/M1/M2 & 17.4/22.8/23.1 & 
$0.28_{-0.02}^{+0.04}$ & $1.75_{-0.04}^{+0.06}$ & $0.20_{-0.03}^{+0.02}$ & 
$0.95_{-0.05}^{+0.03}$ & $0.26_{+0.01}^{+0.08}$ & 919.5/908 \\
2 & 2003 Nov 25 & M1/M2 & 2.5/2.5 & $0.29_{-0.05}^{+0.06}$ & 
$2.04_{-0.15}^{+0.16}$ & \nodata & $1.44_{-0.18}^{+0.12}$ & \nodata & 63.6/70 \\
3 & 2003 Dec 21 & PN & 7.4 & $0.34_{-0.02}^{+0.02}$ & 
$2.42_{-0.06}^{+0.07}$ & \nodata & $2.18_{-0.10}^{+0.06}$ & \nodata & 346.6/360
 \\
4 & 2003 Dec 23 & PN & 3.2 & $0.34_{-0.03}^{+0.03}$ & 
$2.29_{-0.12}^{+0.13}$ & \nodata & $1.82_{-0.14}^{+0.09}$ & \nodata & 94.9/125
 \\
5 & 2003 Dec 25 & PN & 5.7 & $0.26_{-0.03}^{+0.04}$ & 
$2.16_{-0.13}^{+0.14}$ & \nodata & $1.35_{-0.10}^{+0.09}$ & \nodata & 94.6/103
 \\
6 & 2004 Jan 8 & PN & 6.3 & $0.31_{-0.04}^{+0.06}$ & 
$2.09_{-0.19}^{+0.16}$ & $0.26_{-0.07}^{+0.09}$ & $1.53_{-0.22}^{+0.14}$ & 
$0.30_{-0.01}^{+0.36}$ & 255.4/257 \\
7 & 2004 Jan 17 & PN/M1/M2 & 2.6/6.7/6.8 & $0.33_{-0.02}^{+0.02}$
 & $2.38_{-0.07}^{+0.08}$ & \nodata & $1.78_{-0.10}^{+0.05}$ & \nodata & 
295.9/313 \\
8 & 2004 May 1 & M1/M2 & 7.0/7.4 & $0.34_{-0.09}^{+0.16}$ & 
$1.8_{-0.2}^{+0.2}$ & $0.22_{-0.06}^{+0.09}$ & $1.02_{-0.27}^{+0.06}$ & 
$0.33_{-0.06}^{+0.19}$ & 116.6/113 \\
9 & 2004 Jun 5 & PN & 8.8 & $0.34_{-0.02}^{+0.02}$ & 
$2.39_{-0.05}^{+0.06}$ & \nodata & $3.11_{-0.13}^{+0.05}$ & \nodata & 382.1/398
 \\
10 & 2004 Aug 23 & PN/M1/M2 & 2.6/10.9/11.1 & 
$0.47_{-0.08}^{+0.06}$ & $3.08_{-0.14}^{+0.15}$ & $0.16_{-0.02}^{+0.05}$ & 
$2.2_{-0.8}^{+0.1}$ & $0.9_{-0.2}^{+0.2}$ & 297.1/270 \\
11 & 2004 Nov 23 & M1/M2 & 15.5/15.5 & $0.25_{-0.04}^{+0.04}$ & 
$1.80_{-0.19}^{+0.14}$ & $0.30_{-0.06}^{+0.08}$ & $1.05_{-0.11}^{+0.12}$ & 
$0.21_{+0.01}^{+0.28}$ & 263.3/276 \\
12 & 2005 Feb 7 & PN/M1/M2 & 6.7/11.7/11.7 & $0.29_{-0.04}^{+0.07}$
 & $1.74_{-0.05}^{+0.11}$ & $0.21_{-0.04}^{+0.03}$ & $1.08_{-0.12}^{+0.05}$ & 
$0.29_{-0.01}^{+0.13}$ & 431.4/426 \\
\noalign{\smallskip} \hline \noalign{\smallskip}
&&&&&NGC 1313 X-2&&&&&\\
\noalign{\smallskip} \hline \noalign{\smallskip}
1 & 2000 Oct 17 & PN & 17.4 & $0.26_{-0.04}^{+0.06}$ & 
$2.12_{-0.10}^{+0.17}$ & $0.24_{-0.07}^{+0.08}$ & $0.34_{-0.05}^{+0.04}$ & 
$0.07_{-0.01}^{+0.09}$ & 230.9/223 \\
2 & 2003 Nov 25 & PN/M1/M2 & 1.0/2.5/2.5 & $0.25_{-0.04}^{+0.05}$
 & $1.81_{-0.12}^{+0.12}$ & \nodata & $0.83_{-0.05}^{+0.07}$ & \nodata & 
115.3/113 \\
3 & 2003 Dec 21 & PN/M1/M2 & 7.4/10.1/10.1 & $0.39_{-0.03}^{+0.14}$
 & $1.86_{-0.05}^{+0.10}$ & $0.12_{-0.01}^{+0.01}$ & $1.41_{-0.17}^{+0.07}$ & 
$0.33_{-0.04}^{+0.26}$ & 608.8/620 \\
4 & 2003 Dec 23 & PN/M1/M2 & 3.2/5.0/5.1 & $0.26_{-0.02}^{+0.02}$
 & $1.71_{-0.05}^{+0.06}$ & \nodata & $1.13_{-0.04}^{+0.05}$ & \nodata & 
316.3/344 \\
5 & 2003 Dec 25 & PN/M1/M2 & 5.7/6.8/7.5 & $0.27_{-0.02}^{+0.02}$
 & $2.22_{-0.07}^{+0.08}$ & \nodata & $0.55_{-0.03}^{+0.02}$ & \nodata & 
304.3/294 \\
6 & 2004 Jan 8 & PN/M1/M2 & 6.3/11.3/12.0 & $0.29_{-0.02}^{+0.02}$
 & $2.39_{-0.07}^{+0.08}$ & \nodata & $0.49_{-0.03}^{+0.02}$ & \nodata & 
299.9/310 \\
7 & 2004 Jan 17 & PN/M1/M2 & 2.6/6.7/6.8 & $0.27_{-0.03}^{+0.03}$
 & $2.34_{-0.11}^{+0.12}$ & \nodata & $0.42_{-0.03}^{+0.03}$ & \nodata & 
138.8/149 \\
8 & 2004 May 1 & M1/M2 & 6.9/7.4 & $0.32_{-0.05}^{+0.05}$ & 
$2.50_{-0.17}^{+0.18}$ & \nodata & $0.38_{-0.05}^{+0.03}$ & \nodata & 71.8/73 \\
9 & 2004 Jun 5 & PN/M1/M2 & 8.8/11.4/11.4 & $0.49_{-0.07}^{+0.08}$
 & $1.93_{-0.05}^{+0.06}$ & $0.12_{-0.01}^{+0.01}$ & $2.25_{-0.31}^{+0.11}$ & 
$0.90_{-0.08}^{+0.20}$ & 760.0/779 \\
10 & 2004 Aug 23 & PN/M1/M2 & 2.6/10.9/11.1 & 
$0.27_{-0.05}^{+0.12}$ & $1.94_{-0.17}^{+0.24}$ & $0.22_{-0.07}^{+0.06}$ & 
$0.36_{-0.07}^{+0.02}$ & $0.11_{-0.01}^{+0.07}$ & 164.3/163 \\
11 & 2004 Nov 23 & PN/M1/M2 & 12.6/15.4/15.4 & 
$0.29_{-0.03}^{+0.07}$ & $2.19_{-0.12}^{+0.14}$ & $0.22_{-0.06}^{+0.05}$ & 
$0.38_{-0.05}^{+0.03}$ & $0.09_{+0.01}^{+0.07}$ & 350.0/365 \\
12 & 2005 Feb 7 & PN/M1/M2 & 6.7/11.7/11.7 & $0.57_{-0.09}^{+0.06}$
 & $1.90_{-0.03}^{+0.05}$ & $0.11_{-0.01}^{+0.01}$ & $3.2_{-0.7}^{+0.3}$ & 
$2.0_{-0.4}^{+0.3}$ & 728.8/663 \\
\enddata

\tablecomments{Instruments: data from which instrument, PN, MOS1(M1) or MOS2(M2), are used; Exposure: clean exposures for corresponding instruments after background flares excluded; $n_{\rm H}$: column density along the line of sight; $\Gamma$: power-law photon index; $kT$: inner disk temperature of the multicolor disk component, which is unavailable when the the disk component is less than 99\% of confidence level; $L_{\rm X}$: 0.3--10 keV intrinsic luminosity; $L_{\rm disk}$: 0.3--10 keV disk luminosity; $\chi^2$/dof: $\chi^2$ and degree of freedom for the best-fit model; All errors are in 90\% confidence level.}

\end{deluxetable}

\end{document}